\documentclass[prl,superscriptaddress,twocolumn,epsf,nofootnotebib,showpacs]{revtex4-1}

\usepackage{amsfonts,amssymb,amsmath}
\usepackage{mathrsfs}
\usepackage{color}
\usepackage{graphicx}
\usepackage{dcolumn}
\usepackage{bm}

\newcommand{\C}[1]{{\mathcal #1}}
\newcommand{\F}[1]{{\mathfrak #1}}

\newcommand{\BF}[1]{{\mathbf #1}}

\newcommand{\half}{\frac 12}

\newcommand{\Slash}[1]{{\ooalign{\hfil#1\hfil\crcr\raise.167ex\hbox{/}}}}

\begin{document}

\title{Gauss-Bonnet Chern-Simons gravitational wave leptogenesis}

\author{Shinsuke Kawai}
\email{kawai@skku.edu}
\affiliation{Department of Physics, Sungkyunkwan University,
Suwon 16419, Republic of Korea}

\author{Jinsu Kim}
\email{kimjinsu@kias.re.kr}
\affiliation{Quantum Universe Center, Korea Institute for Advanced Study, Seoul 02455, Republic of Korea}

\date{\today}

\begin{abstract}
The gravitational Chern-Simons term coupled to an evolving axion is known to generate lepton number through the gravitational anomaly.
We examine this leptogenesis scenario in the presence of the Gauss-Bonnet term over and above the gravitational Chern-Simons term.
We find that the lepton production can be exponentially enhanced.
The Gauss-Bonnet term creates CP-violating instability of gravitational waves
that may appear transiently after inflation, and during the period of instability elliptically polarized gravitational waves are exponentially amplified at sub-horizon scales.
This instability does not affect the spectrum of the cosmic microwave background as it occurs at much shorter length scales.
In a typical scenario based on {\em natural inflation}, the observed baryon asymmetry of the Universe corresponds to the UV cutoff scale at $10^{14-16}$ GeV.
\end{abstract}

\pacs{
98.80.Cq, 
04.30.Nk, 
11.30.Er, 
11.30.Fs 
}
\keywords{Inflation}
\maketitle

The Universe is observed to be baryon-asymmetric and the origin of this asymmetry remains as an unsettled problem of particle physics.
Half a century ago, Sakharov \cite{Sakharov:1967dj}
elucidated conditions for baryogenesis:
baryon number violation, C and CP violation, and departure from thermal equilibrium.
Realization of these conditions within the electroweak phase transition of the Standard Model (SM) is known to be problematic \cite{Huet:1994jb}, and this difficulty signals the necessity for a theory beyond the SM. 
Popular scenarios of baryogenesis, such as the GUT baryogenesis \cite{Kolb:1979qa} and the Fukugita-Yanagida scenario of leptogenesis \cite{Fukugita:1986hr},
typically assume new physics at energy scales of $10^{14-16}$ GeV.
In general, these exquisite scenarios are difficult to test experimentally since the new physics is at such high energy.
The model of leptogenesis proposed by Alexander, Peskin and Sheikh-Jabbari
\cite{Alexander:2004us} uses elliptically polarized gravitational waves as the source of the lepton asymmetry. 
In their model, the lepton number is generated during inflation by the gravitational anomaly \cite{Delbourgo:1972xb,Eguchi:1976db,AlvarezGaume:1983ig}
\begin{align}\label{eqn:anomaly}
  \nabla_\mu J_{\ell}^\mu=\frac{3}{16\pi^2}R\widetilde R,
\end{align}
where $J_\ell^\mu$ is the lepton number current and
$R\widetilde R\equiv\half\epsilon^{\mu\nu\rho\sigma}R_{\mu\nu\kappa\lambda}R^{\kappa\lambda}{}_{\rho\sigma}$ 
is the gravitational Chern-Simons (gCS) term.
The lepton number is then converted into baryon number via the sphaleron processes
\cite{Kuzmin:1985mm,Khlebnikov:1988sr}.
As pointed out by Lue, Wang and Kamionkowski \cite{Lue:1998mq},
vestiges of elliptically polarized gravitational waves may be found in the cosmic microwave background (CMB).
Moreover, the primordial gravitational waves that may be directly responsible for leptogenesis must be present today as gravitational wave background, and
in view of the remarkable progress of gravitational wave astronomy
\cite{Abbott:2016blz} one may hope for direct detection in the future.
In the original proposal \cite{Alexander:2004us}, the elliptically polarized gravitational waves are sourced by an evolving axion field $\varphi$ (identified as the inflaton) coupled to $R\widetilde R$.
We point out in this Letter that the lepton production due to the anomaly \eqref{eqn:anomaly} can be exponentially enhanced if the axion is coupled also to the Gauss-Bonnet (GB) term 
$R_{\rm GB}^2\equiv R^2-4R_{\mu\nu}R^{\mu\nu}+R^{\mu\nu\rho\sigma}R_{\mu\nu\rho\sigma}$.
This term is in the same order (quadratic curvature) and has similar topological nature as the gCS term.
The occurrence of the GB term is also natural from the viewpoint of inflationary effective field theory \cite{Weinberg:2008hq}.

\begin{figure}[t]
  \includegraphics[width=85mm]{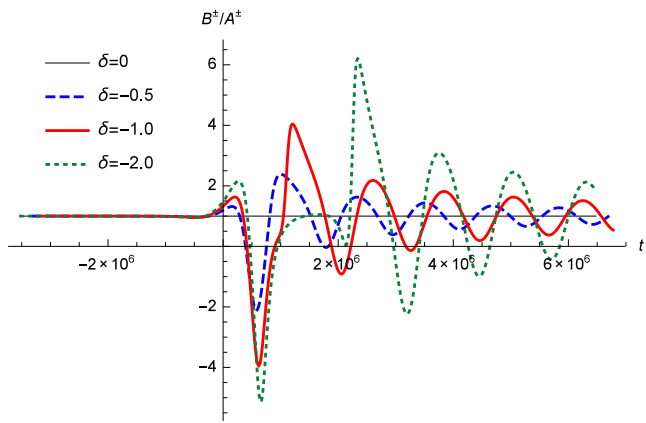}
  \caption{\label{fig:BoverA}
Time evolution of $B^\pm/A^\pm$ for {\em natural inflation}, with the potential \eqref{eqn:V} and the axion-GB coupling \eqref{eqn:xi}.
The strength of the coupling is varied $\delta=0$, $-0.5$, $-1.0$, $-2.0$.
The axion decay constant is $f=8.7 M_{\rm P}$, and $\Lambda$ is normalized by the scalar power spectrum $A_s=2.207\times 10^{-9}$ at $N_e=60$.
The origin of time $t=0$ is at the end of the slow roll.
The gCS coupling is $\vartheta=0$ in this plot and the unit of $M_{\rm P}=1$ is used.}
\end{figure}

Our observation is as follows.
Consider the action
\begin{align}\label{eqn:action}
	S=&\int d^4 x\sqrt{-g}\Big\{\frac{M_{\rm P}^2}{2} R
	-\half\partial_\mu\varphi\partial^\mu\varphi-V(\varphi)\crcr
	&-\frac{1}{16}\xi(\varphi)R_{\rm GB}^2
	+\frac{1}{16}\vartheta(\varphi)R\widetilde R
	\Big\},
\end{align}
where $M_{\rm P}
=2.44\times 10^{18}$ GeV is the reduced Planck mass and 
$\varphi$ an axion field that drives inflation.
CPT invariance requires $\vartheta(\varphi)$ to be an odd function and $V(\varphi)$, $\xi(\varphi)$ to be even functions of $\varphi$.
%
Assuming spatial homogeneity of the axion and the FRW metric with tensor perturbation
$ds^2=-dt^2+a^2 [e^h]_{ij}dx^idx^j$, where
$[e^h]_{ij}\equiv\delta_{ij}+h_{ij}+\half h_{ik}h^k{}_j+\cdots$,
the equations for the helicity components of the transverse-traceless tensor $h_{ij}$ are 
\cite{Kawai:1998ab,Satoh:2007gn,Satoh:2008ck,Satoh:2010ep}
\begin{align}\label{eqn:gwEoM}
  \ddot h^\pm_{\BF k}
  +\left(3H+\frac{\dot A^\pm}{A^\pm}\right)\dot h^\pm_{\BF k}
  +\left(\frac ka\right)^2\frac {B^\pm}{A^\pm} h^\pm_{\BF k}=0,
\end{align}
with
\begin{align}\label{eqn:AandB}
  A^\pm\equiv 1-\frac{H\dot\xi}{2M_{\rm P}^2} \mp\frac{k}{a} \frac{\dot\vartheta}{2M_{\rm P}^2},\quad
  B^\pm\equiv 1-\frac{\ddot\xi}{2M_{\rm P}^2} \mp\frac{k}{a} \frac{\dot\vartheta}{2M_{\rm P}^2}.
\end{align}
Dots denote cosmic time derivatives.
As we will see in an example below (Fig.\ref{fig:BoverA}), under reasonable assumptions $B^\pm/A^\pm$ in the last term of \eqref{eqn:gwEoM} can be negative.
This gives exponential amplification of the gravitational waves, and if the gravitational waves are elliptically polarized it leads to enhancement of the lepton asymmetry through the anomaly \eqref{eqn:anomaly}.
If the GB term is absent, $B^\pm/A^\pm=1$ identically and the coefficient of the last term in \eqref{eqn:gwEoM} is always positive; therefore such an enhancement never occurs.

{\em Benchmark scenario.}---
As a concrete example we focus on the axion potential
\begin{align}\label{eqn:V}
  V(\varphi)=\Lambda^4\left(1+\cos\frac{\varphi}{f}\right)
\end{align}
and the coupling to the GB term
\begin{align}\label{eqn:xi}
  \xi(\varphi)=-\delta\ln\left[2 e^{\varphi} \eta^4(ie^{\varphi})\right],
\end{align}
where $\eta(\tau)\equiv q^{1/24}\prod_{n=1}^\infty (1-q^n)$ is the Dedekind function and $q\equiv e^{2\pi i\tau}$.
Here and below $M_{\rm P}$ is suppressed unless indicated explicitly.
The coupling of the form \eqref{eqn:xi} arises as threshold corrections in certain heterotic and type II superstring compactifications
\cite{Antoniadis:1992sa,Antoniadis:1992rq,Antoniadis:1993ze} (see also \cite{Florakis:2016aoi}).
The parameter $\delta$ can take both signs depending on the number of supermultiplets. 
Due to the property $\eta(-1/\tau)=\sqrt{-i\tau}\; \eta(\tau)$,
the function $\xi(\varphi)$ is an even function as required.
We choose a linear function $\vartheta(\varphi)=\gamma\varphi$ for the gCS coupling.
The GB and gCS terms break the shift symmetry of the axion. 
In general, global symmetries are expected to be broken by quantum gravity effects (discussed e.g. in \cite{Banks:2010zn}); the coupling of the axion to the higher curvature may be regarded as such effects.

The model \eqref{eqn:V} is known as {\em natural inflation} 
\cite{Freese:1990rb,Adams:1992bn}.
In the absence of the higher curvature terms, it is 2$\sigma$-consistent with the Planck 2015 data for $N_e=60$ e-folds (the fit becomes worse for smaller $N_e$) \cite{Ade:2015lrj}.
While the gCS term does not modify the background inflaton dynamics, the GB term does.
The Friedmann and Klein-Gordon equations are modified and are obtained from the action \eqref{eqn:action} as
\begin{align}\label{eqn:bgEoM}
  &3M_{\rm P}^2 H^2=\half\dot\varphi^2+V+\frac 32 H^3\xi_{,\varphi}\dot\varphi,\crcr
  &\ddot\varphi+3H\dot\varphi+V_{,\varphi}+\frac 32 H^2(\dot H+H^2)\xi_{,\varphi}=0.
\end{align}
In fact, a scalar field coupled to the GB term can drive accelerated expansion of the Universe even without a potential term, and the viability of such alternative inflation models has been a focus of much attention \cite{Antoniadis:1993jc,Kawai:1999pw,Kawai:1999xn,Satoh:2007gn,Satoh:2008ck,Satoh:2010ep}.
Here we take a conservative view that the higher curvature terms are corrections to the Einstein gravity, and discuss a scenario in which the effects of such terms are minor during the slow roll.
This view is natural in our model, as the GB and gCS terms are topological terms that become trivial if the axion field is held constant.
Those terms turn out to be significant when the axion becomes dynamical, that is, towards the end of inflation.

The coupling to the GB term modifies the axion dynamics as follows.
When $|\delta|$ is small the dynamics is same as the original {\em natural inflation} model: slow roll in the region $0<\varphi<\pi f$, followed by damped oscillations about the potential minimum at $\varphi=\pi f$.
Reheating of the Universe takes place in the oscillating phase.
Nonzero $\delta$ creates a de Sitter fixed point, in the region $0<\varphi<\pi f$ if $\delta>0$ and $\pi f<\varphi<2\pi f$ if $\delta<0$.
As a consequence, when $\delta$ is larger than a certain positive value the axion stops at the fixed point before reaching the potential minimum; thus inflation does not terminate.
If $\delta$ is smaller than a certain negative value the axion will overshoot the potential minimum and never come back, which means there is no reheating.
As a benchmark we choose $f=8.7 M_{\rm P}$, which in the absence of the GB term gives the best-fit scalar spectral index 
$n_s=0.9652$ of the Planck 2015 data (TT, TE, EE+lowP) \cite{Ade:2015lrj} for $N_e=60$.
The parameter $\Lambda$ is fixed by the amplitude of the scalar power spectrum  $A_s=2.207\times 10^{-9}$ (Planck 2015, same as above).
With this parameter choice the range of $\delta$ for successful inflation, graceful exit and reheating turns out to be $-2.5 \lesssim\delta\lesssim 0.2$.
The tensor/scalar ratio at the CMB scale ($N_e=60$, say) tends to be suppressed
when $\delta$ is negative.
The observation thus favors $\delta<0$.

Fig.\ref{fig:BoverA} shows the behavior of $B^\pm/A^\pm$ in this model,
with $\vartheta=0$ and $\delta$ varied as $0$, $-0.5$, $-1.0$, $-2.0$.
The origin of time $t=0$ is at the end of inflation when the slow roll parameter $\epsilon_H\equiv -\dot H/H^2$ becomes unity.
For moderate nonzero values of $\delta$ the ratio $B^\pm/A^\pm$ is seen to become negative temporarily after the slow roll.
During this period the amplitudes of $h_{\BF k}^\pm$ grow exponentially.
In this model $B^\pm$ changes the sign while $A^\pm$ stays positive.
As the last term of \eqref{eqn:gwEoM} is proportional to $k^2/a^2$, only the gravitational waves at sub-horizon scales are amplified;
the fluctuations at the CMB scale are not affected as they are already way outside the horizon.
When $\dot\vartheta\neq 0$, the growth rates of the left/right-polarized waves differ. 
This asymmetry is however secondary to the asymmetry of the initial states as long as the energy is sub-Planckian ($k/a\lesssim M_{\rm P}$) and $\dot\vartheta$ is not large in the Planck unit.

\begin{figure*}[t]
\includegraphics[width=59mm]{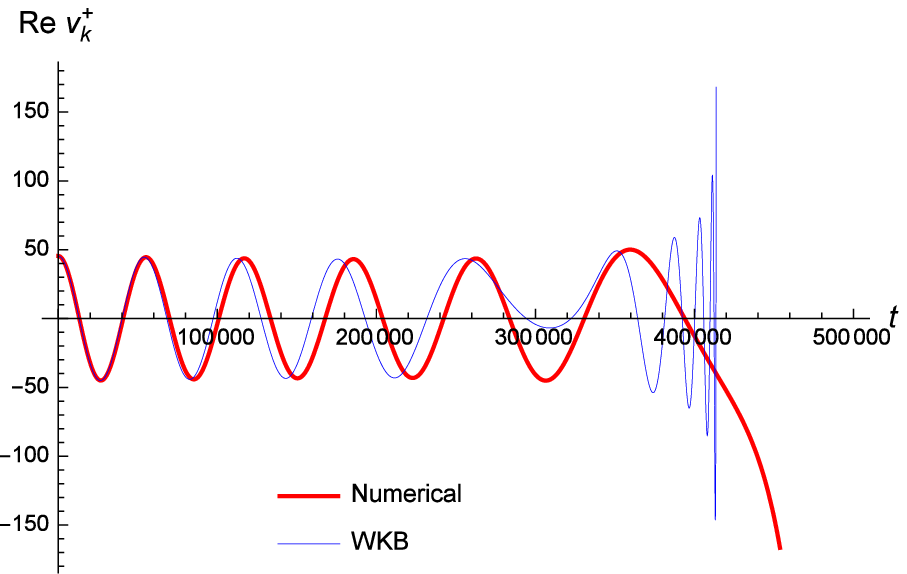}
\includegraphics[width=59mm]{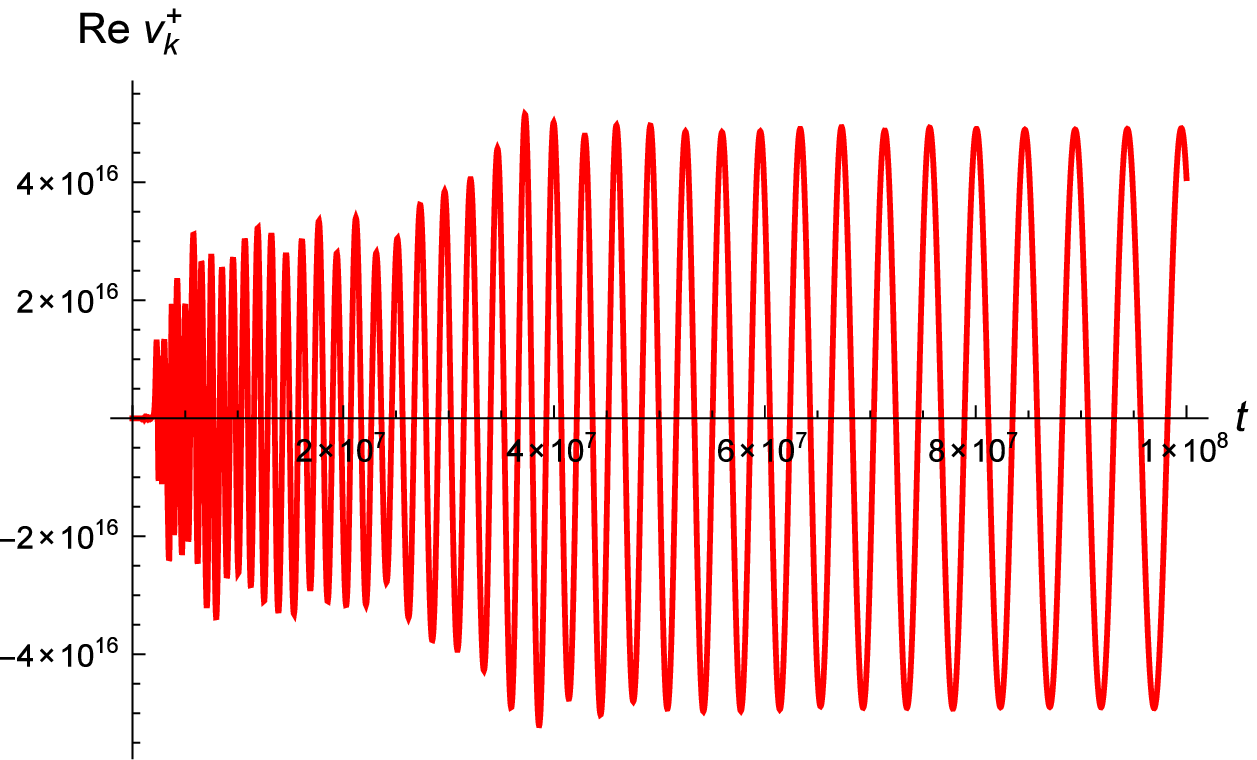}
\includegraphics[width=59mm]{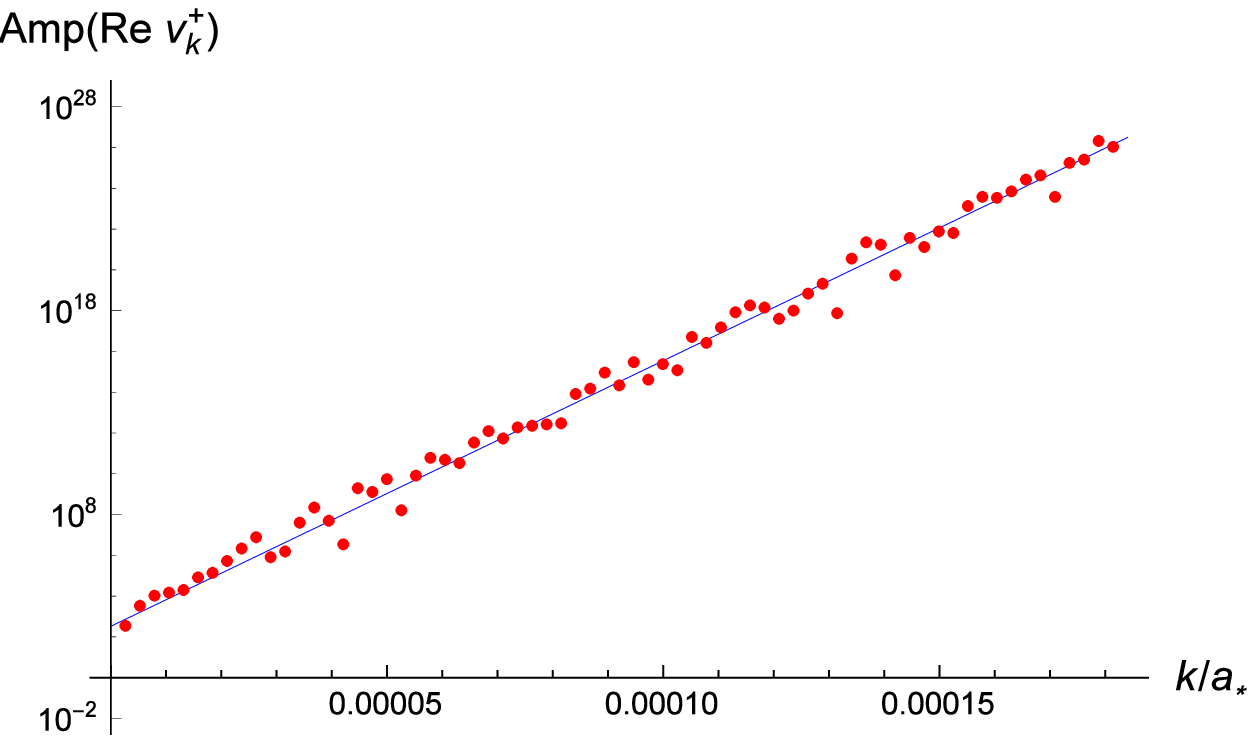}
\caption{\label{fig:modefn}
Evolution of the normalized mode functions $v^\pm_{\BF k}$ of the gravitational waves.
The inflaton dynamics is the same as the $\delta=-1$ case of Fig.\ref{fig:BoverA}.
The gCS coupling is $\gamma=1$ here, and the unit of $M_{\rm P}=1$ is used.
{\em Left} and {\em center}:
the figures show the real part of the left-helicity mode $v^+_{\BF k}$ with $k/a_*=20\, m$ ($m=\Lambda^2/f$ is the inflaton mass scale and $a_*$ is the scale factor at $t=0$).
The oscillatory behavior matches well with the WKB solution (the thin blue curve), until $t\approx 413632$ when $B^\pm/A^\pm$ becomes negative and the amplitude starts to grow (left).
After experiencing the exponential growth, the amplitude of the oscillations becomes stable again (center). 
{\em Right panel}: the stabilized amplitude of ${\rm Re}\; v^+_{\BF k}$ at large enough $t$ is plotted against the wavenumber, in the range $m\leq k/a_*\leq 35\; m$.
The straight line is ${\rm Amp}({\rm Re}\; v^+_{\BF k})\propto \exp(\Xi k/a_*)$, with $\Xi=3.0\times 10^5$.
}
\end{figure*}

{\em Amplification of CP-violating gravitational waves.}---
The key element of this leptogenesis scenario is the CP-violating tensor mode fluctuations that are generated near the end of inflation and amplified at the onset of reheating.
As the physics is similar to the tachyonic resonance scenario of reheating \cite{Felder:2000hj,Felder:2001kt}, we use the technologies developed 
in \cite{Dolgov:1989us,Traschen:1990sw,Shtanov:1994ce,Kofman:1994rk,Kofman:1997yn} and apply them to gravitons.

The transverse-traceless tensor $h_{ij}$ contains two propagating degrees of freedom.
For our purposes it is convenient to use the complex dyad $e_i$, $\overline e_j$ to decompose it in a helicity basis,
\begin{align}\label{eqn:hij}
  h_{ij}=\int\frac{d^3 k}{(2\pi)^3}\left(
  h^+_{\BF k} e_ie_j+h^-_{\BF k}\overline e_i\overline e_j\right)
  e^{i{\BF k}\cdot{\BF x}}.
\end{align}
The dyad is normalized as $\delta^{ij}e_i\overline e_j=1$ and satisfies 
$\delta^{ij}e_ie_j=\delta^{ij}\overline e_i\overline e_j=0$, $q^i e_i=q^i\overline e_i=0$,
$\epsilon^{ijk} q_ie_j\overline e_k=-iq$ for a given 3-momentum $q^i$ and totally antisymmetric $\epsilon^{ijk}$ ($\epsilon^{123}=1$).
The helicity components 
are expanded,
\begin{align}
  h^\pm_{\BF k}
  =a^\pm_{\BF k}{\F h}^\pm_{\BF k}+a^\mp{}^\dag_{-{\BF k}}({\F h}^\mp_{-{\BF k}})^*,
\end{align}
using the operators satisfying the quantization conditions
  $[a^\pm_{\BF k}, a^\pm{}^\dag_{\BF l}]=(2\pi)^3\delta^3({\BF k}-{\BF l})$ and 
  $[a^\pm_{\BF k}, a^\mp{}^\dag_{\BF l}]=0$.
The relation $h^-_{\BF k}=(h^+_{-{\BF k}})^\dag$ follows from the reality of the metric.
The mode functions ${\F h}^\pm_{\BF k}$ satisfy the same equations as the classical counterparts \eqref{eqn:gwEoM}.
In the presence of the gCS term, the modes ${\F h}^+_{\BF k}$ and ${\F h}^-_{\BF k}$ evolve differently.
With
$F^\pm_{\BF k}\equiv a \sqrt{A^\pm}$
the normalized mode functions are defined
\begin{align}
  v^\pm_{\BF k}\equiv \frac{M_{\rm P}}{2}F^\pm_{\BF k}{\F h}^\pm_{\BF k}
\end{align}
and satisfy the flat-space Klein-Gordon equations
\begin{align}\label{eqn:HO}
  (v^\pm_{\BF k})''+(\omega^\pm_{\BF k})^2 v^\pm_{\BF k}=0.
\end{align}
A prime denotes differentiation with respect to the conformal time $\eta=\int dt/a$. 
The angular frequencies are
\begin{align}\label{eqn:omega}
  \omega^\pm_{\BF k}\equiv\sqrt{k^2\frac{B^\pm}{A^\pm}-\frac{(F^\pm_{\BF k})''}{F^\pm_{\BF k}}}.
\end{align}
During slow roll inflation, the flat space limit 
can be taken deep inside the horizon and solutions to \eqref{eqn:HO} are found in the WKB approximation.
Assuming the Bunch-Davies vacuum we choose the positive frequency mode
\begin{align}\label{eqn:WKB}
  v^\pm_{\BF k}\simeq\frac{e^{-i\omega^\pm_{\BF k}\eta}}{2\sqrt{\omega^\pm_{\BF k}}}.
\end{align}
These solutions are also valid near the end of inflation when the slow roll conditions are marginally satisfied.

The initial conditions for the gravitational waves are set by \eqref{eqn:WKB}  at time $t=0$ when the slow roll parameter $\epsilon_H= -\dot H/H^2$ reaches unity, and the subsequent evolution is studied by solving \eqref{eqn:HO} numerically.
In the study of reheating, it is known that resonant decay of an inflaton in an expanding universe exhibits stochasticity \cite{Dolgov:1989us,Traschen:1990sw,Shtanov:1994ce,Kofman:1994rk,Kofman:1997yn}.
In contrast, the dynamics of gravitational waves in our scenario turns out to be relatively simple, presumably because
the instability dies out after a couple of inflaton oscillations.
By numerics we obtain the following results:
(i) from the initial time $t=0$ up until $B^\pm/A^\pm$ becomes negative, the WKB solutions \eqref{eqn:WKB} are in good agreement with numerical solutions;
(ii) during the instability period ($B^\pm/A^\pm<0$) the amplitudes of the mode functions grow exponentially;
(iii) when $B^\pm/A^\pm$ becomes positive again, the normalized mode functions $v^\pm_{\BF k}$ start to oscillate with stable amplitudes 
${\rm Amp}(v^\pm_{\BF k})$;
(iv) ${\rm Amp}(v^\pm_{\BF k})$ exhibit exponential dependence on the wavenumber
\begin{align}\label{eqn:kdep}
  {\rm Amp}(v^\pm_{\BF k})\propto e^{\Xi k/a_*},
\end{align}
where $a_*$ is the scale factor at time $t=0$.
The exponent is found to be $\Xi\sim3.0\times 10^5$ in the benchmark scenario with 
$\delta=-1$ and $\gamma=1$. 
Fig.\ref{fig:modefn} shows the results for (the real part of) the left-helicity mode $v^+_{\BF k}$.
The behavior of the right-helicity mode $v^-_{\BF k}$ is nearly identical.
We see from $B^\pm$ of \eqref{eqn:AandB} that for positive $\gamma$, 
$v^+_{\BF k}$ enters the instability earlier and exits later than $v^-_{\BF k}$.
This gives enhancement of the asymmetry, but its effects on the final amplitudes are minor in this benchmark model.

The $k$-dependence of the amplification factor \eqref{eqn:kdep} is understood as follows.
During the instability $B^\pm/A^\pm<0$, 
$(\widetilde\omega^\pm_{\BF k})^2\equiv -(\omega^\pm_{\BF k})^2
\approx k^2 |B^\pm/A^\pm|>0$ since $(F^\pm_{\BF k})''/F^\pm_{\BF k}$ in \eqref{eqn:omega} is subdominant.
Then \eqref{eqn:HO} become
$(v^\pm_{\BF k})''-(\widetilde\omega^\pm_{\BF k})^2 v^\pm_{\BF k}=0$,
which have growing solutions
$v^\pm_{\BF k}\propto e^{\widetilde\omega^\pm_{\BF k}\eta}$.
Denoting the time scale of instability as
$\Delta t=a\Delta\eta\approx a_*\Delta\eta$,
the growth of the mode functions is 
$e^{\widetilde\omega^\pm_{\BF k}\Delta\eta}\approx e^{k |B^\pm/A^\pm|^{1/2}\Delta t /a_*}$.
Comparing this with \eqref{eqn:kdep}, we identify
$\Xi\approx |B^\pm/A^\pm|^{1/2} \Delta t$.
The instability is driven by the background inflaton oscillations and its time scale is determined by the inflaton mass, $\Delta t\sim 1/m$.
Fig.\ref{fig:BoverA} shows $|B^\pm/A^\pm|$ is ${\C O}(1)$ in our benchmark model and thus, $\Xi$ is essentially the inverse of the inflaton mass.
For {\em natural inflation},
$m=\Lambda^2/f\sim 5\times 10^{-6} M_{\rm P}$, giving
$\Xi\sim\Delta t\sim 1/m\sim 2\times 10^5 M_{\rm P}^{-1}$.
This agrees with our numerical estimate up to a factor of ${\C O}(1)$.

{\em Lepton asymmetry.}---
Evaluation of lepton number amounts to finding the expectation value of $R\widetilde R$ in the Bunch-Davies vacuum. 
Classically, the gCS term is 
\begin{align}\label{eqn:GCS}
  R\widetilde R= -\frac{2\epsilon^{ijk}}{a^4}\left(
  h''_{j\ell}h'^\ell{}_{k|i}
  -h'_{j\ell}{}^{|m}h^\ell{}_{k}{}_{|im}
  +h'_{j}{}^{m|\ell}h_{k\ell|mi}
  \right),
\end{align}
where the vertical strokes denote spatial derivatives and higher order terms are neglected.
Substituting \eqref{eqn:hij} into \eqref{eqn:GCS} and using the Weyl ordering of operators
we find
\begin{align}
\langle R\widetilde R\rangle
=\frac{1}{2\pi^2 a^4}\int k^3 dk \frac{d\Delta_{\BF k}}{d\eta},
\end{align}
with
  $
  \Delta_{\BF k}\equiv
  {\F h}^+_{\BF k}{}'{\F h}^+_{\BF k}{}^*{}'
  -{\F h}^-_{\BF k}{}'{\F h}^-_{\BF k}{}^*{}'
  -k^2({\F h}^+_{\BF k}{\F h}^+_{\BF k}{}^*
  -{\F h}^-_{\BF k}{\F h}^-_{\BF k}{}^*)
  $.
Lepton number $L$ is integration of the current \eqref{eqn:anomaly}.
Using the comoving volume $\Omega\equiv\int dx^3$, lepton number density is
\begin{align}\label{eqn:nL}
  n_L
  =\frac{L}{a^3\Omega}
  =\frac{3}{32\pi^4}\frac{1}{a^3}\int_{k_{\rm IR}}^{k_{\rm UV}} k^3dk\Delta_{\BF k}(\eta).
\end{align}
The upper bound of the $k$-integral is given by the cutoff scale of the theory $\mu=k_{\rm UV}/a$.
As the anomaly \eqref{eqn:anomaly} arises from the absence of right-handed neutrinos in the SM, a natural choice of $\mu$ may be the seesaw scale.
Alternatively, $\mu$ can be the GUT scale or the Planck scale if the lepton sector is to be the same up to these scales.
The lower bound $k_{\rm IR}$ may be chosen at the horizon scale $k_{\rm IR}=aH$, or simply zero.
The lepton number is sensitive to the choice of $k_{\rm UV}$, but insensitive to $k_{\rm IR}$.

Let us first ignore the amplification due to the GB term and evaluate the lepton asymmetry using the solutions in the sub-horizon limit \eqref{eqn:WKB}.
In slow roll approximation (which is marginally valid) and to the leading order in $k$, 
\begin{align}\label{eqn:Deltak}
  \Delta_{\BF k}\simeq -\frac{H^2\dot\vartheta}{a M_{\rm P}^4}.
\end{align}
Note that $k$-dependence drops in the leading order. 
Then $k$-integration in \eqref{eqn:nL} gives lepton number density
\begin{align}
  n_L\simeq -\frac{3}{128\pi^4}\frac{H^6\dot\vartheta}{M_{\rm P}^4}\left(
  \frac{\mu}{H}\right)^4.
\end{align}
This is essentially the same result as Alexander et al. \cite{Alexander:2004us}.
Assuming reheating to be instantaneous, the reheating temperature $T$ is related to $H$ at the onset of reheating and the entropy density is written
$s=(2\pi^2/45)g_*T^3\simeq 2.3\times g_*^{1/4} (M_{\rm P} H)^{3/2}$.
Using the SM degrees of freedom $g_*\sim 100$, the lepton yield becomes
\begin{align}\label{eqn:YLwoGB}
  \frac{n_L}{s}\simeq 10^{-4}\times \frac{\dot\vartheta}{H}\left(\frac{H}{M_{\rm P}}\right)^\frac 32 \left(\frac{\mu}{M_{\rm P}}\right)^4.
\end{align}
This is to be compared with $n_L/s\approx 2.4\times 10^{-10}$
given by the baryon density from the Planck experiment \cite{Ade:2015xua} and the sphaleron conversion factor ${n_B}/{n_L}=-{28}/{79}$.
The Planck constraints on the tensor/scalar ratio $r\lesssim 0.1$ set an upper bound on 
$H$ 
at the time of the horizon exit of the CMB scale, giving 
$({H}/{M_{\rm P}})^\frac 32\lesssim 10^{-7}$.
The Hubble parameter at later time is smaller and the constraints become tighter.
Using the slow roll parameter $\epsilon_V\equiv (M_{\rm P}V_{,\varphi}/V)^2/2$
one may write $\dot\vartheta/H\sim \gamma M_{\rm P}\sqrt{2\epsilon_V}$, which is not much larger than, say, $10^3$ in realistic string theoretical setup, and Alexander et al. \cite{Alexander:2004us} concluded that the cutoff around the Planck scale would be appropriate. 
Dilution due to cosmic expansion may put more severe constraints and
the effects of renormalisation may further suppress lepton production \cite{Fischler:2007tj}.
Of course, instantaneous reheating is a strong assumption and uncertainties of the reheating scenario may change the prediction.
If entropy production is extremely inefficient, $n_L/s$ would be larger than \eqref{eqn:YLwoGB} and the constraints can be relaxed.

In our scenario the gravitational waves are amplified during the instability phase by an exponential factor $\exp(\Xi k/a)$.
Accordingly, \eqref{eqn:Deltak} is modified to
\begin{align}
  \Delta_{\BF k}\simeq -\frac{H^2\dot\vartheta}{a M_{\rm P}^4}e^{2\Xi k/a},
\end{align}
and neglecting subleading terms \eqref{eqn:nL} gives
\begin{align}
  n_L\simeq -\frac{3}{64\pi^4}\frac{H^2\dot\vartheta\mu^3}{M_{\rm P}^4\Xi}e^{2\Xi\mu}.
\end{align}
Assuming instantaneous reheating, the lepton yield is
\begin{align}
  \frac{n_L}{s}
  \simeq 10^{-4}\times 
  \frac{e^{2\Xi\mu}}{M_{\rm P}\Xi}
  \frac{\dot\vartheta}{H}
  \left(\frac{H}{M_{\rm P}}\right)^\frac 32
  \left(\frac{\mu}{M_{\rm P}}\right)^3,
\end{align}
where $H$ and $\dot\vartheta$ are to be evaluated at the end of inflation.
This is our main result.
Let us modestly take $\dot\vartheta/H\sim {\C O}(1)$.
In {\em natural inflation}, the Hubble parameter at the end of inflation is $H\sim 10^{-6} M_{\rm P}$; hence $(H/M_{\rm P})^{3/2}\sim 10^{-9}$.
As discussed above, $\Xi$ is roughly the inverse of the inflaton mass, 
$\Xi\sim 1/m\sim 2\times 10^5/M_{\rm P}$.
Then, $\mu\sim 25 m\sim 3\times 10^{14}$ GeV gives $n_L/s\sim 10^{-10}$.
Taking the dilution of leptons and other ambiguities into account, the cutoff scale of $\mu\sim 10^{14-16}$ GeV would be reasonable.


%
%

{\em Acknowledgements.}---
We acknowledge helpful conversations with Alejandro Ibarra.
This work was supported in part by the National Research Foundation (Korea) Grant-in-Aid for Scientific Research No. NRF-2015R1D1A1A01061507 (S.K.).



\begin{thebibliography}{37}%
\makeatletter
\providecommand \@ifxundefined [1]{%
 \@ifx{#1\undefined}
}%
\providecommand \@ifnum [1]{%
 \ifnum #1\expandafter \@firstoftwo
 \else \expandafter \@secondoftwo
 \fi
}%
\providecommand \@ifx [1]{%
 \ifx #1\expandafter \@firstoftwo
 \else \expandafter \@secondoftwo
 \fi
}%
\providecommand \natexlab [1]{#1}%
\providecommand \enquote  [1]{``#1''}%
\providecommand \bibnamefont  [1]{#1}%
\providecommand \bibfnamefont [1]{#1}%
\providecommand \citenamefont [1]{#1}%
\providecommand \href@noop [0]{\@secondoftwo}%
\providecommand \href [0]{\begingroup \@sanitize@url \@href}%
\providecommand \@href[1]{\@@startlink{#1}\@@href}%
\providecommand \@@href[1]{\endgroup#1\@@endlink}%
\providecommand \@sanitize@url [0]{\catcode `\\12\catcode `\$12\catcode
  `\&12\catcode `\#12\catcode `\^12\catcode `\_12\catcode `\%12\relax}%
\providecommand \@@startlink[1]{}%
\providecommand \@@endlink[0]{}%
\providecommand \url  [0]{\begingroup\@sanitize@url \@url }%
\providecommand \@url [1]{\endgroup\@href {#1}{\urlprefix }}%
\providecommand \urlprefix  [0]{URL }%
\providecommand \Eprint [0]{\href }%
\providecommand \doibase [0]{http://dx.doi.org/}%
\providecommand \selectlanguage [0]{\@gobble}%
\providecommand \bibinfo  [0]{\@secondoftwo}%
\providecommand \bibfield  [0]{\@secondoftwo}%
\providecommand \translation [1]{[#1]}%
\providecommand \BibitemOpen [0]{}%
\providecommand \bibitemStop [0]{}%
\providecommand \bibitemNoStop [0]{.\EOS\space}%
\providecommand \EOS [0]{\spacefactor3000\relax}%
\providecommand \BibitemShut  [1]{\csname bibitem#1\endcsname}%
\let\auto@bib@innerbib\@empty
\bibitem [{\citenamefont {Sakharov}(1967)}]{Sakharov:1967dj}%
  \BibitemOpen
  \bibfield  {author} {\bibinfo {author} {\bibfnamefont {A.~D.}\ \bibnamefont
  {Sakharov}},\ }\href {\doibase 10.1070/PU1991v034n05ABEH002497} {\bibfield
  {journal} {\bibinfo  {journal} {Pisma Zh. Eksp. Teor. Fiz.}\ }\textbf
  {\bibinfo {volume} {5}},\ \bibinfo {pages} {32} (\bibinfo {year} {1967})},\
  \bibinfo {note} {[Usp. Fiz. Nauk161,61(1991)]}\BibitemShut {NoStop}%
\bibitem [{\citenamefont {Huet}\ and\ \citenamefont
  {Sather}(1995)}]{Huet:1994jb}%
  \BibitemOpen
  \bibfield  {author} {\bibinfo {author} {\bibfnamefont {P.}~\bibnamefont
  {Huet}}\ and\ \bibinfo {author} {\bibfnamefont {E.}~\bibnamefont {Sather}},\
  }\href {\doibase 10.1103/PhysRevD.51.379} {\bibfield  {journal} {\bibinfo
  {journal} {Phys. Rev.}\ }\textbf {\bibinfo {volume} {D51}},\ \bibinfo {pages}
  {379} (\bibinfo {year} {1995})},\ \Eprint
  {http://arxiv.org/abs/hep-ph/9404302} {arXiv:hep-ph/9404302 [hep-ph]}
  \BibitemShut {NoStop}%
\bibitem [{\citenamefont {Kolb}\ and\ \citenamefont
  {Wolfram}(1980)}]{Kolb:1979qa}%
  \BibitemOpen
  \bibfield  {author} {\bibinfo {author} {\bibfnamefont {E.~W.}\ \bibnamefont
  {Kolb}}\ and\ \bibinfo {author} {\bibfnamefont {S.}~\bibnamefont {Wolfram}},\
  }\href {\doibase 10.1016/0550-3213(80)90167-4} {\bibfield  {journal}
  {\bibinfo  {journal} {Nucl.Phys.}\ }\textbf {\bibinfo {volume} {B172}},\
  \bibinfo {pages} {224} (\bibinfo {year} {1980})}\BibitemShut {NoStop}%
\bibitem [{\citenamefont {Fukugita}\ and\ \citenamefont
  {Yanagida}(1986)}]{Fukugita:1986hr}%
  \BibitemOpen
  \bibfield  {author} {\bibinfo {author} {\bibfnamefont {M.}~\bibnamefont
  {Fukugita}}\ and\ \bibinfo {author} {\bibfnamefont {T.}~\bibnamefont
  {Yanagida}},\ }\href {\doibase 10.1016/0370-2693(86)91126-3} {\bibfield
  {journal} {\bibinfo  {journal} {Phys.Lett.}\ }\textbf {\bibinfo {volume}
  {B174}},\ \bibinfo {pages} {45} (\bibinfo {year} {1986})}\BibitemShut
  {NoStop}%
\bibitem [{\citenamefont {Alexander}\ \emph {et~al.}(2006)\citenamefont
  {Alexander}, \citenamefont {Peskin},\ and\ \citenamefont
  {Sheikh-Jabbari}}]{Alexander:2004us}%
  \BibitemOpen
  \bibfield  {author} {\bibinfo {author} {\bibfnamefont {S.~H.-S.}\
  \bibnamefont {Alexander}}, \bibinfo {author} {\bibfnamefont {M.~E.}\
  \bibnamefont {Peskin}}, \ and\ \bibinfo {author} {\bibfnamefont {M.~M.}\
  \bibnamefont {Sheikh-Jabbari}},\ }\href {\doibase
  10.1103/PhysRevLett.96.081301} {\bibfield  {journal} {\bibinfo  {journal}
  {Phys.Rev.Lett.}\ }\textbf {\bibinfo {volume} {96}},\ \bibinfo {pages}
  {081301} (\bibinfo {year} {2006})},\ \Eprint
  {http://arxiv.org/abs/hep-th/0403069} {arXiv:hep-th/0403069 [hep-th]}
  \BibitemShut {NoStop}%
\bibitem [{\citenamefont {Delbourgo}\ and\ \citenamefont
  {Salam}(1972)}]{Delbourgo:1972xb}%
  \BibitemOpen
  \bibfield  {author} {\bibinfo {author} {\bibfnamefont {R.}~\bibnamefont
  {Delbourgo}}\ and\ \bibinfo {author} {\bibfnamefont {A.}~\bibnamefont
  {Salam}},\ }\href {\doibase 10.1016/0370-2693(72)90825-8} {\bibfield
  {journal} {\bibinfo  {journal} {Phys. Lett.}\ }\textbf {\bibinfo {volume}
  {B40}},\ \bibinfo {pages} {381} (\bibinfo {year} {1972})}\BibitemShut
  {NoStop}%
\bibitem [{\citenamefont {Eguchi}\ and\ \citenamefont
  {Freund}(1976)}]{Eguchi:1976db}%
  \BibitemOpen
  \bibfield  {author} {\bibinfo {author} {\bibfnamefont {T.}~\bibnamefont
  {Eguchi}}\ and\ \bibinfo {author} {\bibfnamefont {P.~G.~O.}\ \bibnamefont
  {Freund}},\ }\href {\doibase 10.1103/PhysRevLett.37.1251} {\bibfield
  {journal} {\bibinfo  {journal} {Phys. Rev. Lett.}\ }\textbf {\bibinfo
  {volume} {37}},\ \bibinfo {pages} {1251} (\bibinfo {year}
  {1976})}\BibitemShut {NoStop}%
\bibitem [{\citenamefont {Alvarez-Gaume}\ and\ \citenamefont
  {Witten}(1984)}]{AlvarezGaume:1983ig}%
  \BibitemOpen
  \bibfield  {author} {\bibinfo {author} {\bibfnamefont {L.}~\bibnamefont
  {Alvarez-Gaume}}\ and\ \bibinfo {author} {\bibfnamefont {E.}~\bibnamefont
  {Witten}},\ }\href {\doibase 10.1016/0550-3213(84)90066-X} {\bibfield
  {journal} {\bibinfo  {journal} {Nucl.Phys.}\ }\textbf {\bibinfo {volume}
  {B234}},\ \bibinfo {pages} {269} (\bibinfo {year} {1984})}\BibitemShut
  {NoStop}%
\bibitem [{\citenamefont {Kuzmin}\ \emph {et~al.}(1985)\citenamefont {Kuzmin},
  \citenamefont {Rubakov},\ and\ \citenamefont {Shaposhnikov}}]{Kuzmin:1985mm}%
  \BibitemOpen
  \bibfield  {author} {\bibinfo {author} {\bibfnamefont {V.}~\bibnamefont
  {Kuzmin}}, \bibinfo {author} {\bibfnamefont {V.}~\bibnamefont {Rubakov}}, \
  and\ \bibinfo {author} {\bibfnamefont {M.}~\bibnamefont {Shaposhnikov}},\
  }\href {\doibase 10.1016/0370-2693(85)91028-7} {\bibfield  {journal}
  {\bibinfo  {journal} {Phys.Lett.}\ }\textbf {\bibinfo {volume} {B155}},\
  \bibinfo {pages} {36} (\bibinfo {year} {1985})}\BibitemShut {NoStop}%
\bibitem [{\citenamefont {Khlebnikov}\ and\ \citenamefont
  {Shaposhnikov}(1988)}]{Khlebnikov:1988sr}%
  \BibitemOpen
  \bibfield  {author} {\bibinfo {author} {\bibfnamefont {S.~Y.}\ \bibnamefont
  {Khlebnikov}}\ and\ \bibinfo {author} {\bibfnamefont {M.}~\bibnamefont
  {Shaposhnikov}},\ }\href {\doibase 10.1016/0550-3213(88)90133-2} {\bibfield
  {journal} {\bibinfo  {journal} {Nucl.Phys.}\ }\textbf {\bibinfo {volume}
  {B308}},\ \bibinfo {pages} {885} (\bibinfo {year} {1988})}\BibitemShut
  {NoStop}%
\bibitem [{\citenamefont {Lue}\ \emph {et~al.}(1999)\citenamefont {Lue},
  \citenamefont {Wang},\ and\ \citenamefont {Kamionkowski}}]{Lue:1998mq}%
  \BibitemOpen
  \bibfield  {author} {\bibinfo {author} {\bibfnamefont {A.}~\bibnamefont
  {Lue}}, \bibinfo {author} {\bibfnamefont {L.-M.}\ \bibnamefont {Wang}}, \
  and\ \bibinfo {author} {\bibfnamefont {M.}~\bibnamefont {Kamionkowski}},\
  }\href {\doibase 10.1103/PhysRevLett.83.1506} {\bibfield  {journal} {\bibinfo
   {journal} {Phys.Rev.Lett.}\ }\textbf {\bibinfo {volume} {83}},\ \bibinfo
  {pages} {1506} (\bibinfo {year} {1999})},\ \Eprint
  {http://arxiv.org/abs/astro-ph/9812088} {arXiv:astro-ph/9812088 [astro-ph]}
  \BibitemShut {NoStop}%
\bibitem [{\citenamefont {Abbott}\ \emph {et~al.}(2016)\citenamefont {Abbott}
  \emph {et~al.}}]{Abbott:2016blz}%
  \BibitemOpen
  \bibfield  {author} {\bibinfo {author} {\bibfnamefont {B.~P.}\ \bibnamefont
  {Abbott}} \emph {et~al.} (\bibinfo {collaboration} {Virgo, LIGO
  Scientific}),\ }\href {\doibase 10.1103/PhysRevLett.116.061102} {\bibfield
  {journal} {\bibinfo  {journal} {Phys. Rev. Lett.}\ }\textbf {\bibinfo
  {volume} {116}},\ \bibinfo {pages} {061102} (\bibinfo {year} {2016})},\
  \Eprint {http://arxiv.org/abs/1602.03837} {arXiv:1602.03837 [gr-qc]}
  \BibitemShut {NoStop}%
\bibitem [{\citenamefont {Weinberg}(2008)}]{Weinberg:2008hq}%
  \BibitemOpen
  \bibfield  {author} {\bibinfo {author} {\bibfnamefont {S.}~\bibnamefont
  {Weinberg}},\ }\href {\doibase 10.1103/PhysRevD.77.123541} {\bibfield
  {journal} {\bibinfo  {journal} {Phys.Rev.}\ }\textbf {\bibinfo {volume}
  {D77}},\ \bibinfo {pages} {123541} (\bibinfo {year} {2008})},\ \Eprint
  {http://arxiv.org/abs/0804.4291} {arXiv:0804.4291 [hep-th]} \BibitemShut
  {NoStop}%
\bibitem [{\citenamefont {Kawai}\ \emph {et~al.}(1998)\citenamefont {Kawai},
  \citenamefont {Sakagami},\ and\ \citenamefont {Soda}}]{Kawai:1998ab}%
  \BibitemOpen
  \bibfield  {author} {\bibinfo {author} {\bibfnamefont {S.}~\bibnamefont
  {Kawai}}, \bibinfo {author} {\bibfnamefont {M.-A.}\ \bibnamefont {Sakagami}},
  \ and\ \bibinfo {author} {\bibfnamefont {J.}~\bibnamefont {Soda}},\ }\href
  {\doibase 10.1016/S0370-2693(98)00925-3} {\bibfield  {journal} {\bibinfo
  {journal} {Phys.Lett.}\ }\textbf {\bibinfo {volume} {B437}},\ \bibinfo
  {pages} {284} (\bibinfo {year} {1998})},\ \Eprint
  {http://arxiv.org/abs/gr-qc/9802033} {arXiv:gr-qc/9802033 [gr-qc]}
  \BibitemShut {NoStop}%
\bibitem [{\citenamefont {Satoh}\ \emph {et~al.}(2008)\citenamefont {Satoh},
  \citenamefont {Kanno},\ and\ \citenamefont {Soda}}]{Satoh:2007gn}%
  \BibitemOpen
  \bibfield  {author} {\bibinfo {author} {\bibfnamefont {M.}~\bibnamefont
  {Satoh}}, \bibinfo {author} {\bibfnamefont {S.}~\bibnamefont {Kanno}}, \ and\
  \bibinfo {author} {\bibfnamefont {J.}~\bibnamefont {Soda}},\ }\href {\doibase
  10.1103/PhysRevD.77.023526} {\bibfield  {journal} {\bibinfo  {journal} {Phys.
  Rev.}\ }\textbf {\bibinfo {volume} {D77}},\ \bibinfo {pages} {023526}
  (\bibinfo {year} {2008})},\ \Eprint {http://arxiv.org/abs/0706.3585}
  {arXiv:0706.3585 [astro-ph]} \BibitemShut {NoStop}%
\bibitem [{\citenamefont {Satoh}\ and\ \citenamefont
  {Soda}(2008)}]{Satoh:2008ck}%
  \BibitemOpen
  \bibfield  {author} {\bibinfo {author} {\bibfnamefont {M.}~\bibnamefont
  {Satoh}}\ and\ \bibinfo {author} {\bibfnamefont {J.}~\bibnamefont {Soda}},\
  }\href {\doibase 10.1088/1475-7516/2008/09/019} {\bibfield  {journal}
  {\bibinfo  {journal} {JCAP}\ }\textbf {\bibinfo {volume} {0809}},\ \bibinfo
  {pages} {019} (\bibinfo {year} {2008})},\ \Eprint
  {http://arxiv.org/abs/0806.4594} {arXiv:0806.4594 [astro-ph]} \BibitemShut
  {NoStop}%
\bibitem [{\citenamefont {Satoh}(2010)}]{Satoh:2010ep}%
  \BibitemOpen
  \bibfield  {author} {\bibinfo {author} {\bibfnamefont {M.}~\bibnamefont
  {Satoh}},\ }\href {\doibase 10.1088/1475-7516/2010/11/024} {\bibfield
  {journal} {\bibinfo  {journal} {JCAP}\ }\textbf {\bibinfo {volume} {1011}},\
  \bibinfo {pages} {024} (\bibinfo {year} {2010})},\ \Eprint
  {http://arxiv.org/abs/1008.2724} {arXiv:1008.2724 [astro-ph.CO]} \BibitemShut
  {NoStop}%
\bibitem [{\citenamefont {Antoniadis}\ \emph
  {et~al.}(1992{\natexlab{a}})\citenamefont {Antoniadis}, \citenamefont
  {Gava},\ and\ \citenamefont {Narain}}]{Antoniadis:1992sa}%
  \BibitemOpen
  \bibfield  {author} {\bibinfo {author} {\bibfnamefont {I.}~\bibnamefont
  {Antoniadis}}, \bibinfo {author} {\bibfnamefont {E.}~\bibnamefont {Gava}}, \
  and\ \bibinfo {author} {\bibfnamefont {K.~S.}\ \bibnamefont {Narain}},\
  }\href {\doibase 10.1016/0370-2693(92)90009-S} {\bibfield  {journal}
  {\bibinfo  {journal} {Phys. Lett.}\ }\textbf {\bibinfo {volume} {B283}},\
  \bibinfo {pages} {209} (\bibinfo {year} {1992}{\natexlab{a}})},\ \Eprint
  {http://arxiv.org/abs/hep-th/9203071} {arXiv:hep-th/9203071 [hep-th]}
  \BibitemShut {NoStop}%
\bibitem [{\citenamefont {Antoniadis}\ \emph
  {et~al.}(1992{\natexlab{b}})\citenamefont {Antoniadis}, \citenamefont
  {Gava},\ and\ \citenamefont {Narain}}]{Antoniadis:1992rq}%
  \BibitemOpen
  \bibfield  {author} {\bibinfo {author} {\bibfnamefont {I.}~\bibnamefont
  {Antoniadis}}, \bibinfo {author} {\bibfnamefont {E.}~\bibnamefont {Gava}}, \
  and\ \bibinfo {author} {\bibfnamefont {K.~S.}\ \bibnamefont {Narain}},\
  }\href {\doibase 10.1016/0550-3213(92)90672-X} {\bibfield  {journal}
  {\bibinfo  {journal} {Nucl. Phys.}\ }\textbf {\bibinfo {volume} {B383}},\
  \bibinfo {pages} {93} (\bibinfo {year} {1992}{\natexlab{b}})},\ \Eprint
  {http://arxiv.org/abs/hep-th/9204030} {arXiv:hep-th/9204030 [hep-th]}
  \BibitemShut {NoStop}%
\bibitem [{\citenamefont {Antoniadis}\ \emph
  {et~al.}(1994{\natexlab{a}})\citenamefont {Antoniadis}, \citenamefont {Gava},
  \citenamefont {Narain},\ and\ \citenamefont {Taylor}}]{Antoniadis:1993ze}%
  \BibitemOpen
  \bibfield  {author} {\bibinfo {author} {\bibfnamefont {I.}~\bibnamefont
  {Antoniadis}}, \bibinfo {author} {\bibfnamefont {E.}~\bibnamefont {Gava}},
  \bibinfo {author} {\bibfnamefont {K.~S.}\ \bibnamefont {Narain}}, \ and\
  \bibinfo {author} {\bibfnamefont {T.~R.}\ \bibnamefont {Taylor}},\ }\href
  {\doibase 10.1016/0550-3213(94)90617-3} {\bibfield  {journal} {\bibinfo
  {journal} {Nucl. Phys.}\ }\textbf {\bibinfo {volume} {B413}},\ \bibinfo
  {pages} {162} (\bibinfo {year} {1994}{\natexlab{a}})},\ \Eprint
  {http://arxiv.org/abs/hep-th/9307158} {arXiv:hep-th/9307158 [hep-th]}
  \BibitemShut {NoStop}%
\bibitem [{\citenamefont {Florakis}(2017)}]{Florakis:2016aoi}%
  \BibitemOpen
  \bibfield  {author} {\bibinfo {author} {\bibfnamefont {I.}~\bibnamefont
  {Florakis}},\ }\href {\doibase 10.1016/j.nuclphysb.2017.01.016} {\bibfield
  {journal} {\bibinfo  {journal} {Nucl. Phys.}\ }\textbf {\bibinfo {volume}
  {B916}},\ \bibinfo {pages} {484} (\bibinfo {year} {2017})},\ \Eprint
  {http://arxiv.org/abs/1611.10323} {arXiv:1611.10323 [hep-th]} \BibitemShut
  {NoStop}%
\bibitem [{\citenamefont {Banks}\ and\ \citenamefont
  {Seiberg}(2011)}]{Banks:2010zn}%
  \BibitemOpen
  \bibfield  {author} {\bibinfo {author} {\bibfnamefont {T.}~\bibnamefont
  {Banks}}\ and\ \bibinfo {author} {\bibfnamefont {N.}~\bibnamefont
  {Seiberg}},\ }\href {\doibase 10.1103/PhysRevD.83.084019} {\bibfield
  {journal} {\bibinfo  {journal} {Phys. Rev.}\ }\textbf {\bibinfo {volume}
  {D83}},\ \bibinfo {pages} {084019} (\bibinfo {year} {2011})},\ \Eprint
  {http://arxiv.org/abs/1011.5120} {arXiv:1011.5120 [hep-th]} \BibitemShut
  {NoStop}%
\bibitem [{\citenamefont {Freese}\ \emph {et~al.}(1990)\citenamefont {Freese},
  \citenamefont {Frieman},\ and\ \citenamefont {Olinto}}]{Freese:1990rb}%
  \BibitemOpen
  \bibfield  {author} {\bibinfo {author} {\bibfnamefont {K.}~\bibnamefont
  {Freese}}, \bibinfo {author} {\bibfnamefont {J.~A.}\ \bibnamefont {Frieman}},
  \ and\ \bibinfo {author} {\bibfnamefont {A.~V.}\ \bibnamefont {Olinto}},\
  }\href {\doibase 10.1103/PhysRevLett.65.3233} {\bibfield  {journal} {\bibinfo
   {journal} {Phys. Rev. Lett.}\ }\textbf {\bibinfo {volume} {65}},\ \bibinfo
  {pages} {3233} (\bibinfo {year} {1990})}\BibitemShut {NoStop}%
\bibitem [{\citenamefont {Adams}\ \emph {et~al.}(1993)\citenamefont {Adams},
  \citenamefont {Bond}, \citenamefont {Freese}, \citenamefont {Frieman},\ and\
  \citenamefont {Olinto}}]{Adams:1992bn}%
  \BibitemOpen
  \bibfield  {author} {\bibinfo {author} {\bibfnamefont {F.~C.}\ \bibnamefont
  {Adams}}, \bibinfo {author} {\bibfnamefont {J.~R.}\ \bibnamefont {Bond}},
  \bibinfo {author} {\bibfnamefont {K.}~\bibnamefont {Freese}}, \bibinfo
  {author} {\bibfnamefont {J.~A.}\ \bibnamefont {Frieman}}, \ and\ \bibinfo
  {author} {\bibfnamefont {A.~V.}\ \bibnamefont {Olinto}},\ }\href {\doibase
  10.1103/PhysRevD.47.426} {\bibfield  {journal} {\bibinfo  {journal} {Phys.
  Rev.}\ }\textbf {\bibinfo {volume} {D47}},\ \bibinfo {pages} {426} (\bibinfo
  {year} {1993})},\ \Eprint {http://arxiv.org/abs/hep-ph/9207245}
  {arXiv:hep-ph/9207245 [hep-ph]} \BibitemShut {NoStop}%
\bibitem [{\citenamefont {Ade}\ \emph {et~al.}(2016{\natexlab{a}})\citenamefont
  {Ade} \emph {et~al.}}]{Ade:2015lrj}%
  \BibitemOpen
  \bibfield  {author} {\bibinfo {author} {\bibfnamefont {P.~A.~R.}\
  \bibnamefont {Ade}} \emph {et~al.} (\bibinfo {collaboration} {Planck}),\
  }\href {\doibase 10.1051/0004-6361/201525898} {\bibfield  {journal} {\bibinfo
   {journal} {Astron. Astrophys.}\ }\textbf {\bibinfo {volume} {594}},\
  \bibinfo {pages} {A20} (\bibinfo {year} {2016}{\natexlab{a}})},\ \Eprint
  {http://arxiv.org/abs/1502.02114} {arXiv:1502.02114 [astro-ph.CO]}
  \BibitemShut {NoStop}%
\bibitem [{\citenamefont {Antoniadis}\ \emph
  {et~al.}(1994{\natexlab{b}})\citenamefont {Antoniadis}, \citenamefont
  {Rizos},\ and\ \citenamefont {Tamvakis}}]{Antoniadis:1993jc}%
  \BibitemOpen
  \bibfield  {author} {\bibinfo {author} {\bibfnamefont {I.}~\bibnamefont
  {Antoniadis}}, \bibinfo {author} {\bibfnamefont {J.}~\bibnamefont {Rizos}}, \
  and\ \bibinfo {author} {\bibfnamefont {K.}~\bibnamefont {Tamvakis}},\ }\href
  {\doibase 10.1016/0550-3213(94)90120-1} {\bibfield  {journal} {\bibinfo
  {journal} {Nucl. Phys.}\ }\textbf {\bibinfo {volume} {B415}},\ \bibinfo
  {pages} {497} (\bibinfo {year} {1994}{\natexlab{b}})},\ \Eprint
  {http://arxiv.org/abs/hep-th/9305025} {arXiv:hep-th/9305025 [hep-th]}
  \BibitemShut {NoStop}%
\bibitem [{\citenamefont {Kawai}\ and\ \citenamefont
  {Soda}(1999{\natexlab{a}})}]{Kawai:1999pw}%
  \BibitemOpen
  \bibfield  {author} {\bibinfo {author} {\bibfnamefont {S.}~\bibnamefont
  {Kawai}}\ and\ \bibinfo {author} {\bibfnamefont {J.}~\bibnamefont {Soda}},\
  }\href {\doibase 10.1016/S0370-2693(99)00736-4} {\bibfield  {journal}
  {\bibinfo  {journal} {Phys. Lett.}\ }\textbf {\bibinfo {volume} {B460}},\
  \bibinfo {pages} {41} (\bibinfo {year} {1999}{\natexlab{a}})},\ \Eprint
  {http://arxiv.org/abs/gr-qc/9903017} {arXiv:gr-qc/9903017 [gr-qc]}
  \BibitemShut {NoStop}%
\bibitem [{\citenamefont {Kawai}\ and\ \citenamefont
  {Soda}(1999{\natexlab{b}})}]{Kawai:1999xn}%
  \BibitemOpen
  \bibfield  {author} {\bibinfo {author} {\bibfnamefont {S.}~\bibnamefont
  {Kawai}}\ and\ \bibinfo {author} {\bibfnamefont {J.}~\bibnamefont {Soda}},\
  }\href@noop {} {\  (\bibinfo {year} {1999}{\natexlab{b}})},\ \Eprint
  {http://arxiv.org/abs/gr-qc/9906046} {arXiv:gr-qc/9906046 [gr-qc]}
  \BibitemShut {NoStop}%
\bibitem [{\citenamefont {Felder}\ \emph
  {et~al.}(2001{\natexlab{a}})\citenamefont {Felder}, \citenamefont
  {Garcia-Bellido}, \citenamefont {Greene}, \citenamefont {Kofman},
  \citenamefont {Linde},\ and\ \citenamefont {Tkachev}}]{Felder:2000hj}%
  \BibitemOpen
  \bibfield  {author} {\bibinfo {author} {\bibfnamefont {G.~N.}\ \bibnamefont
  {Felder}}, \bibinfo {author} {\bibfnamefont {J.}~\bibnamefont
  {Garcia-Bellido}}, \bibinfo {author} {\bibfnamefont {P.~B.}\ \bibnamefont
  {Greene}}, \bibinfo {author} {\bibfnamefont {L.}~\bibnamefont {Kofman}},
  \bibinfo {author} {\bibfnamefont {A.~D.}\ \bibnamefont {Linde}}, \ and\
  \bibinfo {author} {\bibfnamefont {I.}~\bibnamefont {Tkachev}},\ }\href
  {\doibase 10.1103/PhysRevLett.87.011601} {\bibfield  {journal} {\bibinfo
  {journal} {Phys. Rev. Lett.}\ }\textbf {\bibinfo {volume} {87}},\ \bibinfo
  {pages} {011601} (\bibinfo {year} {2001}{\natexlab{a}})},\ \Eprint
  {http://arxiv.org/abs/hep-ph/0012142} {arXiv:hep-ph/0012142 [hep-ph]}
  \BibitemShut {NoStop}%
\bibitem [{\citenamefont {Felder}\ \emph
  {et~al.}(2001{\natexlab{b}})\citenamefont {Felder}, \citenamefont {Kofman},\
  and\ \citenamefont {Linde}}]{Felder:2001kt}%
  \BibitemOpen
  \bibfield  {author} {\bibinfo {author} {\bibfnamefont {G.~N.}\ \bibnamefont
  {Felder}}, \bibinfo {author} {\bibfnamefont {L.}~\bibnamefont {Kofman}}, \
  and\ \bibinfo {author} {\bibfnamefont {A.~D.}\ \bibnamefont {Linde}},\ }\href
  {\doibase 10.1103/PhysRevD.64.123517} {\bibfield  {journal} {\bibinfo
  {journal} {Phys. Rev.}\ }\textbf {\bibinfo {volume} {D64}},\ \bibinfo {pages}
  {123517} (\bibinfo {year} {2001}{\natexlab{b}})},\ \Eprint
  {http://arxiv.org/abs/hep-th/0106179} {arXiv:hep-th/0106179 [hep-th]}
  \BibitemShut {NoStop}%
\bibitem [{\citenamefont {Dolgov}\ and\ \citenamefont
  {Kirilova}(1990)}]{Dolgov:1989us}%
  \BibitemOpen
  \bibfield  {author} {\bibinfo {author} {\bibfnamefont {A.~D.}\ \bibnamefont
  {Dolgov}}\ and\ \bibinfo {author} {\bibfnamefont {D.~P.}\ \bibnamefont
  {Kirilova}},\ }\href@noop {} {\bibfield  {journal} {\bibinfo  {journal} {Sov.
  J. Nucl. Phys.}\ }\textbf {\bibinfo {volume} {51}},\ \bibinfo {pages} {172}
  (\bibinfo {year} {1990})},\ \bibinfo {note} {[Yad.
  Fiz.51,273(1990)]}\BibitemShut {NoStop}%
\bibitem [{\citenamefont {Traschen}\ and\ \citenamefont
  {Brandenberger}(1990)}]{Traschen:1990sw}%
  \BibitemOpen
  \bibfield  {author} {\bibinfo {author} {\bibfnamefont {J.~H.}\ \bibnamefont
  {Traschen}}\ and\ \bibinfo {author} {\bibfnamefont {R.~H.}\ \bibnamefont
  {Brandenberger}},\ }\href {\doibase 10.1103/PhysRevD.42.2491} {\bibfield
  {journal} {\bibinfo  {journal} {Phys. Rev.}\ }\textbf {\bibinfo {volume}
  {D42}},\ \bibinfo {pages} {2491} (\bibinfo {year} {1990})}\BibitemShut
  {NoStop}%
\bibitem [{\citenamefont {Shtanov}\ \emph {et~al.}(1995)\citenamefont
  {Shtanov}, \citenamefont {Traschen},\ and\ \citenamefont
  {Brandenberger}}]{Shtanov:1994ce}%
  \BibitemOpen
  \bibfield  {author} {\bibinfo {author} {\bibfnamefont {Y.}~\bibnamefont
  {Shtanov}}, \bibinfo {author} {\bibfnamefont {J.~H.}\ \bibnamefont
  {Traschen}}, \ and\ \bibinfo {author} {\bibfnamefont {R.~H.}\ \bibnamefont
  {Brandenberger}},\ }\href {\doibase 10.1103/PhysRevD.51.5438} {\bibfield
  {journal} {\bibinfo  {journal} {Phys. Rev.}\ }\textbf {\bibinfo {volume}
  {D51}},\ \bibinfo {pages} {5438} (\bibinfo {year} {1995})},\ \Eprint
  {http://arxiv.org/abs/hep-ph/9407247} {arXiv:hep-ph/9407247 [hep-ph]}
  \BibitemShut {NoStop}%
\bibitem [{\citenamefont {Kofman}\ \emph {et~al.}(1994)\citenamefont {Kofman},
  \citenamefont {Linde},\ and\ \citenamefont {Starobinsky}}]{Kofman:1994rk}%
  \BibitemOpen
  \bibfield  {author} {\bibinfo {author} {\bibfnamefont {L.}~\bibnamefont
  {Kofman}}, \bibinfo {author} {\bibfnamefont {A.~D.}\ \bibnamefont {Linde}}, \
  and\ \bibinfo {author} {\bibfnamefont {A.~A.}\ \bibnamefont {Starobinsky}},\
  }\href {\doibase 10.1103/PhysRevLett.73.3195} {\bibfield  {journal} {\bibinfo
   {journal} {Phys.Rev.Lett.}\ }\textbf {\bibinfo {volume} {73}},\ \bibinfo
  {pages} {3195} (\bibinfo {year} {1994})},\ \Eprint
  {http://arxiv.org/abs/hep-th/9405187} {arXiv:hep-th/9405187 [hep-th]}
  \BibitemShut {NoStop}%
\bibitem [{\citenamefont {Kofman}\ \emph {et~al.}(1997)\citenamefont {Kofman},
  \citenamefont {Linde},\ and\ \citenamefont {Starobinsky}}]{Kofman:1997yn}%
  \BibitemOpen
  \bibfield  {author} {\bibinfo {author} {\bibfnamefont {L.}~\bibnamefont
  {Kofman}}, \bibinfo {author} {\bibfnamefont {A.~D.}\ \bibnamefont {Linde}}, \
  and\ \bibinfo {author} {\bibfnamefont {A.~A.}\ \bibnamefont {Starobinsky}},\
  }\href {\doibase 10.1103/PhysRevD.56.3258} {\bibfield  {journal} {\bibinfo
  {journal} {Phys.Rev.}\ }\textbf {\bibinfo {volume} {D56}},\ \bibinfo {pages}
  {3258} (\bibinfo {year} {1997})},\ \Eprint
  {http://arxiv.org/abs/hep-ph/9704452} {arXiv:hep-ph/9704452 [hep-ph]}
  \BibitemShut {NoStop}%
\bibitem [{\citenamefont {Ade}\ \emph {et~al.}(2016{\natexlab{b}})\citenamefont
  {Ade} \emph {et~al.}}]{Ade:2015xua}%
  \BibitemOpen
  \bibfield  {author} {\bibinfo {author} {\bibfnamefont {P.~A.~R.}\
  \bibnamefont {Ade}} \emph {et~al.} (\bibinfo {collaboration} {Planck}),\
  }\href {\doibase 10.1051/0004-6361/201525830} {\bibfield  {journal} {\bibinfo
   {journal} {Astron. Astrophys.}\ }\textbf {\bibinfo {volume} {594}},\
  \bibinfo {pages} {A13} (\bibinfo {year} {2016}{\natexlab{b}})},\ \Eprint
  {http://arxiv.org/abs/1502.01589} {arXiv:1502.01589 [astro-ph.CO]}
  \BibitemShut {NoStop}%
\bibitem [{\citenamefont {Fischler}\ and\ \citenamefont
  {Paban}(2007)}]{Fischler:2007tj}%
  \BibitemOpen
  \bibfield  {author} {\bibinfo {author} {\bibfnamefont {W.}~\bibnamefont
  {Fischler}}\ and\ \bibinfo {author} {\bibfnamefont {S.}~\bibnamefont
  {Paban}},\ }\href {\doibase 10.1088/1126-6708/2007/10/066} {\bibfield
  {journal} {\bibinfo  {journal} {JHEP}\ }\textbf {\bibinfo {volume} {0710}},\
  \bibinfo {pages} {066} (\bibinfo {year} {2007})},\ \Eprint
  {http://arxiv.org/abs/0708.3828} {arXiv:0708.3828 [hep-th]} \BibitemShut
  {NoStop}%
\end{thebibliography}
%


%
%

\end{document}